\documentclass[preprint,aps,nofootinbib]{revtex4}
\usepackage{graphicx}
\usepackage{amsmath}
\usepackage{amsfonts}
\usepackage{amssymb}
\usepackage{color}%
\providecommand{\U}[1]{\protect\rule{.1in}{.1in}}
\definecolor{lightgray}{rgb}{.7,.7,.7}

\definecolor{red}{rgb}{1,0,0}

\definecolor{blue}{rgb}{0,0,1}

\definecolor{purple}{rgb}{0.6,0.1,0.7}

\newcommand{\f}{\begin{equation}}
\newcommand{\ff}{\end{equation}}
\newcommand{\fa}{\begin{eqnarray}}
\newcommand{\ffa}{\end{eqnarray}}

\begin{document}
\title{(m,n)-type holographic dark energy models}
\author{Yi Ling $^{1,2}$}
\email{lingy@ihep.ac.cn}
\author{Wen-Jian Pan $^{2,1}$}
\email{wjpan_zhgkxy@163.com}
\affiliation{$^1$Institute of High
Energy Physics, Chinese Academy of Sciences, Beijing 100049,
China\\ $^2$ Center for Relativistic Astrophysics and High Energy
Physics, Department of Physics, Nanchang University, 330031, China
}

\begin{abstract}
We construct $(m,n)$-type holographic dark energy models at a
phenomenological level, which can be viewed as a generalization of
agegraphic models with the conformal-like age as the holographic
characteristic size. For some values of $(m,n)$ the holographic
dark energy can automatically evolve across $\omega=-1$ into a
phantom phase even without introducing an interaction between the
dark energy and background matter. Our construction is also
applicable to the holographic dark energy with generalized future
event horizon as the characteristic size. Finally, we address
the issue on the stability of our model and show that they are
generally stable under the scalar perturbation.
\end{abstract}
\maketitle

\section{Introduction}
Recent cosmological observations have disclosed the current
accelerated expansion of the universe driven by the exotic energy
with negative pressure, which is dubbed as dark
energy(DE)\cite{Riess:1998cb,Perlmutter:1998np,Spergel:2003cb,Tegmark:2003ud}.
The dark energy scenario has attracted a great deal of attention
in the last decade. Despite of many efforts in this subject, the
nature of DE is the most mysterious problem in modern cosmology.
The simplest candidate of dark energy is $\Lambda$CDM model, in
which $\omega=-1$ is constant. Although being consistent with all
observations very well, this model undergoes the fine-tuning
problem and the coincidence
problem\cite{Weinberg:1988cp,Weinberg:2000yb}. After this, a lot
of dynamical DE models have been proposed to solve these problems
(for recent reviews we refer to \cite{Copeland:2006wr,Li:2011sd}).
 As a matter of fact, for any dynamical dark energy model it
contains a free parameter $\omega$ to specify, which in first
principle should be derived at a statistical level, like what we
have done for the ordinary matter ($\omega_m=0$) and radiation
($\omega_r={1\over 3}$). Unfortunately, we know little about the
microscopic property of dark energy such that the statistical
mechanics on dark energy is missing. As a result, one needs to
look for some principle to govern the dynamics of the state
parameter of dark energy such that the evolution of the universe
can be uniquely determined. Recently the most popular strategy is
probably applying the holographic
principle\cite{'tHooft:1993gx,'tHooft:1999bw,Susskind:1994vu,Cohen:1998zx}.
Motivated by this principle, one proposed that in a cosmological
setting the total energy of system with size $L$ should not exceed
the mass of a black hole with the same radius, namely
\begin{equation}
L^{3}\Lambda^{4}=L^{3}\rho_{\Lambda}\leq LM_{p}^{2}.
\end{equation}
While saturating this inequality by choosing the largest $L$ it
gives rise to a holographic energy density
\begin{equation}
\rho_{\Lambda}=3c^{2}M_{p}^{2}L^{-2},\label{HDE}
\end{equation}
where $c$ is a dimensionless constant. One usually calls the dark
energy satisfying equation (\ref{HDE}) as holographic dark energy.
Now, the key issue is how to choose the holographic characteristic
scale $L$. During the past years, there are many possible choices
in the literature\cite{Horava:2000tb,Thomas:2002pq,Fischler:1998st,
Bousso:1999xy,Hsu:2004ri,Li:2004rb,Cai:2007us,Wei:2007ty,Gao:2007ep,Huang:2012nz,Wei:2012wt},
in which the holographic scale $L$ can be identified with the
future event horizon\cite{Li:2004rb}, the conformal age of the
universe\cite{Wei:2007ty} or the Ricci scalar of the
universe\cite{Gao:2007ep}. Specially the model taking the
conformal age of the universe is also dubbed as new agegraphic
dark energy model. Recently, Huang and Wu proposed a new
holographic dark energy model with a conformal-like age of the
universe as the scale $L$ in \cite{Huang:2012nz}. This model can
be consistent with the history from the inflation to the current
universe.

Motivated by above progress in this paper we intend to propose a new
type of holographic dark energy model at a phenomenological level
which is characterized by two numbers $(m,n)$. We will demonstrate
that it is quite general to construct a holographic dark energy
model with an age-like scale as the holographic scale $L$. Originally people
intended to propose this scale under the condition that the
corresponding dark energy should be responsible for the
acceleration of the current universe. As a consequence, the
holographic size was previously proposed to be the future event
horizon or the (conformal) age of the universe in turn in
literature. In this sense, all such scales are proposed at a
phenomenological level. In analogy with these conventional models, perhaps the
direct physical motivation of proposing such characteristic scales
in our paper is still obscure, but we have generalized the previous
holographic dark energy models with significant improvements. In addition,
the new introduced parameters $(m,n)$ provide us more space in theory to fit the observational
data. In particular, when $(m,n)$ take some specific numbers all the
agegraphic-like dark energy models previously proposed in the
literature can be recovered. We will investigate the general
features of $(m,n)$-type holographic dark energy models in this
paper.

In addition, the stability of any dark energy model is
always an important issue. Previously the relevant investigations
on the stability of holographic dark energy model and (new)
agegraphic dark energy models have been appeared
in\cite{Myung:2007pn,Kim:2007iv}. In particular, the recent
work in \cite{Li:2008zq} reveals that the traditional holographic
dark energy model is stable, although the perturbation of
holographic dark energy is nonlocal, different from a usual fluid
whose stability was defined by the sound speed square. In this
paper, we will investigate the stability of our $(m,n)$-type
holographic dark energy models and present an affirmative answer
to this issue.

Our paper is organized as follows. In Sec. 2, we construct the
$(m,n)$ type holographic dark energy model with an age-like scale
as the characteristic size and discuss its general properties
during the various epoches of the universe. The interaction
between the dark energy and the dark (background) matter$(DM)$ is
discussed in Sec. 3 and the coincidence problem is addressed.
In Sec. 4 we briefly remark that our construction is also
applicable to the models with generalized future event horizon as
the holographic size in the same spirit. In Sec. 5 we
shall discuss the issue of the stability of our model.
 The conclusions and discussions are given in Sec. 6.

\section{$(m,n)$-type holographic dark energy models}
We start with the standard Friedmann equations in which DE and
 background constituent with constant of state $\omega_{i}$, are assumed to be
independent, without interaction between them.
\begin{equation}
3M_{p}^{2}H^{2}=\rho_{\Lambda}+\rho_{i},\label{f1}%
\end{equation}%
\begin{equation}
\dot{\rho}_{\Lambda}+3H\rho_{\Lambda}(1+\omega)=0,\label{i1}%
\end{equation}%
\begin{equation}
\dot{\rho}_{i}+3H\rho_{i}(1+{\omega_{i}})=0,\label{i2}%
\end{equation}
where $H=\dot{a}/a$ is the Hubble factor. In particular, $\omega_{i}
= 0$ for pressureless matter, whereas $\omega_{i} = 1/3$ for
radiation. For convenience through this paper we denote the ratio
$\rho_{i}/\rho_{\Lambda}$ by $r$ which is related to
$\Omega_{\Lambda}=\rho_{\Lambda}/(3M_{p}^{2}H^{2})$ by
$1+r=1/\Omega_{\Lambda}$. From the Friedmann equation we easily
obtain a relation between the characteristic size $L$ and the Hubble
factor $H$ as
\begin{equation}
LH=\sqrt{1+r}c.\label{c1}%
\end{equation}
Furthermore, from the equations of conservation, we have
\begin{equation} r'=3({\omega}-{\omega_{i}}) r,\label{der}
\end{equation}
where $r^{\prime}\equiv\dot{r}/H=dr/dlna$. From Eqs.(\ref{f1}),
 (\ref{i1}) and (\ref{i2}) we can work out
\begin{equation}
\dot{H}=-{3\over2}(1+{{{\omega_{i}}r+\omega}\over{1+r}})H^2.\label{h1}
\end{equation}
From Eqs.(\ref{c1}) and (\ref{h1}) one can find
\begin{equation}
2{\frac{L^{\prime}}{L}}=3(1+{\frac{\omega+{\omega_i}r}{1+r}})+{\frac{r^{\prime}}{1+r}}=3(1+\omega).
\label{lo}%
\end{equation}

We point out that the relations derived above are general and
independent of the specific form of the holographic characteristic
scale. Now, we intend to construct a $(m,n)$-type holographic dark
energy model, in which the characteristic scale $L$ is proposed to
be
\begin{equation} L={1\over a^m(t)}\int^t_0a^n(t')dt',\label{s1}
\end{equation}
with $(m,n)$ being a couple of real numbers (at phenomenological
level they need not be integers.). In above definition we
have adopted the scale factor $a(t_0)=1$ for our present
universe. Taking the derivative with respect to $lna$ on both
sides of the equation, we find
\begin{equation} {\frac{L^{\prime}}{L}}=-m+\frac{a^{n-m}}{HL}.
\end{equation}
This relation together with Eq.(\ref{lo}) leads to the equation of
state for $(m,n)$-type holographic dark energy,
\begin{equation} \omega=-1-{2\over 3}m+{2\over 3}\frac{a^{n-m}}{HL}=-1-{2\over 3}m+{2\over 3}\frac{a^{n-m}}{c\sqrt{1+r}}.\label{es1}
\end{equation}
In the absence of the interaction between background matter and dark
energy, Eqs. (\ref{der}) and (\ref{es1}) govern the evolution
of $r$ and $\omega$. Alternatively, one can rewrite the equation of
motion in terms of  $\Omega_{\Lambda}$ as
 \begin{equation}{\Omega^{\prime}_{\Lambda}}={\Omega_{\Lambda}}(1-{\Omega_{\Lambda}})(3+3\omega_i+2m-{{2\sqrt{\Omega_{\Lambda}}}a^{n-m}\over{c}}).\label{w1}
  \end{equation} Next,
we intend to figure out some basic constraints on the values of
$(m,n)$ through the investigation on the general properties of
$(m,n)$-type holographic dark energy during the different epoches of
the universe.

\subsection {Radiation- or Matter-dominated epoch ($a\rightarrow
    0$)}
For a radiation-dominated or matter-dominated epoch, we find the
 Friedmann equation Eq.(\ref{f1}) can be approximately written as,
\begin{equation} \rho_{i} \propto H^2=A^2
a^{-3(1+\omega_{i})},
\end{equation}
where $A$ is a constant and $\omega_{i}$ is the state parameter,
specifically,  $\omega_{i}=1/3$ for radiation and $0$ for matter.
This equation implies that the scale factor evolves as $ a \propto
t^{2\over 3(1+\omega_{i})}.$ Thus, the holographic scale $L$ can be
explicitly integrated out as
\begin{equation} L =\frac{1}{A[n+{3\over 2}(1+\omega_{i})]}
a^{n-m+{3\over 2}(1+\omega_{i})}. \end{equation} This solution leads
to an important relation, implying that the ratio appearing in
Eq.(\ref{es1}) approaches to a constant during the
radiation-dominated or matter-dominated epoch.
\begin{equation} \frac{2a^{n-m}}{3HL}={2\over 3}n+1+\omega_{i}.
\end{equation}
As a result, one can easily find that $\Omega_\Lambda$ during that
epoch evolves as
\begin{equation}
\Omega_\Lambda=(n+{3\over2}+{{3\omega_{i}}\over2})^2c^2a^{2m-2n}.\label{o1}
 \end{equation} It is easy to check that the above equation is consistent with Eq.(\ref{w1}). Moreover, the state parameter
of dark energy is going to a constant, which is
\begin{equation}
\omega={2\over 3}(n-m)+\omega_{i}.
\end{equation}
Obviously, the state parameter depends on the values of $(m,n)$. We
have the following remarks on the constraints on the values of
$(m,n)$.
\begin{itemize}
\item If $n>m$, then $\omega>\omega_{i}\geq 0$. It means $r$
will increase with the expansion of the universe such that the
universe could never exit from a radiation-dominated or
matter-dominated epoch. Thus this case is ruled out and we will
not consider it in next sections.

\item If $n<m$, then $\omega<\omega_{i}$ during the Radiation- or matter-dominated epoch.
In particular, when
    $n-m=-1$, we have $\omega=-{2\over 3}$ and $r\propto a^{-2}$
    for $\omega_i=0$, while $\omega=-{1\over 3}$ and $r\propto a^{-2}$
    for $\omega_i=1/3$. This situation recovers the new agegraphic
    dark energy model \cite{Wei:2007ty} which is $(m,n)=(0,-1)$, and the conformal age-like holographic dark
    energy model which is $(m,n)=(4,3)$\cite{Huang:2012nz}. In
    addition, we notice that in this case as $a\rightarrow 0$,
    the ratio $r$ goes to infinity such that there is no constraint on the
    value of the constant $c$.

\item If $m=n$, then $\omega=\omega_{i}$ and
    $\rho_\Lambda\propto\rho_{i}$. This situation is very
    subtle and previously a similar discussion has been presented for the old agegraphic dark energy
    model which corresponds to the special case with $(m,n)=(0,0)$
    \cite{Cai:2007us}. Since $\omega=\omega_{i}$, the ratio
    between the dark energy and dark matter/radiation would be a
    constant
\begin{equation} r=\frac{1}{c^2[n+{3\over 2}(1+\omega_{i})]^2}-1>0.
\end{equation}
For $\omega_{i}=1/3$, its positivity requires
\begin{equation} c<{1\over m+2}.
\end{equation}
$\omega=\omega_{i}$ implies that dark energy might intend to track
the behavior of the dominated ingredient in the early stage of the
universe, and thus might have the same origin with dark matter. This
potential possibility of unifying dark matter and dark energy is
very interesting. However, to implement this scenario one need
introduce a mechanism to make dark energy deviate from dark matter
and finally the impact of dark energy must be large enough to be
responsible for the acceleration of the universe at late times. We
remark that in the absence of such a mechanism this scenario is hard
to be realized. This difficulty might be overcome by introducing a
suitable interaction between dark energy and dark matter, but here
we leave this issue for further investigation in future.
\end{itemize}
\subsection {Future with $a\gg 1$}
When the interaction between dark energy and dark matter is not
taken into account we only consider the case of $n<m$. When $a\gg
1$, the asymptotic behavior of the universe in future will be
described by the following equations
\begin{equation} \omega=-1-{2\over 3}m,\label{st1}
\end{equation}
\begin{equation} r^{\prime}=-(3+2m)r.
\end{equation}
First of all, if $m\leq -1$, then $\omega>-1/3$. It means that the
universe will not stay in an accelerating phase for ever. An
example discussed in the previous literature is taking the
particle horizon as the holographic characteristic scale, which
corresponds to $m=n=-1$.
\begin{itemize}
    \item If $m>0$, then the holographic dark energy will behave
    like a phantom field.
    \item If $m=0$, then the holographic dark energy will approach
    a cosmological constant. The specific example is the new
    agegraphic dark energy model with $n=-1,m=0$.
    \item If $-1< m<0$, the holographic dark energy
     can drive the universe into an accelerating phase indeed. The key
    point is whether this choice will be consistent with our observational data about the present
    universe.
\end{itemize}

\subsection {Present days}
 The most strict constraints come from the observation data on our
present universe. Here, we only roughly estimate the possible
values for $m$ and $c$. First of all, our current universe has an
accelerating expansion, which requires that
\begin{equation}
1+r_0+3\omega_0<0.\label{h2}%
\end{equation}
We find it leads to
\begin{equation}
m>-1+{r_0\over 2}+\frac{1}{c\sqrt{1+r_0}}.
\end{equation}
Since $c$ is a positive number, if we plug the current $r_0\simeq
{1\over 3}$ into this inequality, then we find a bound for $m$,
which is
\begin{equation} m>-{5\over 6}.
\end{equation}
Conversely, for a given $m$, we find the constant $c$ is
constrained by
\begin{equation} c>{3\sqrt{3}\over 6m+5}.
\end{equation}
If we further require $\omega_0\simeq -1$,  $c$ can be uniquely
fixed by equation, which is $c=1/(m\sqrt{1+r_0})$ ($m>0$ only). In
particular, when $(m,n)=(0,-1)$ and $(m,n)=(4,3)$, our above
estimation is in a good agreement with the results obtained by more
severe constraints from observation
data\cite{Wei:2007xu,Huang:2012gd}. This implies that other types
such as $(m,n)=(1,0),(2,1),(3,2)$ can also fit the data very well.

As a summary, we find the basic constraints on the $(m,n)$-type
holographic dark energy are $n<m$ and $m>-5/6$.

\section{ holographic dark energy with interaction}
Although the nature of both $DM$ and $DE$ still remains a mystery,
the possibility that DE and DM can interact with each other has been
widely discussed
recently\cite{Amendola:1999er,Zimdahl:2001ar,Farrar:2003uw,Gumjudpai:2005ry,Hu:2006ar,CalderaCabral:2008bx,
Wu:2008jt,Sheykhi:2010nv,Jamil:2010xq,Avelino:2012tc,Setare:2006wh,Setare:2007we}. Moreover,
observational signatures on the interaction between dark ones have
been investigated in the probes of the cosmic expansion history with
the use of the SNIa, BAO and CMB shift data
 \cite{Guo:2007zk,He:2008tn,Feng:2008fx}. The interacting dark energy
has also been considered as a possible solution to the coincidence
problem
\cite{Hu:2006ar,TocchiniValentini:2001ty,Cai:2004dk,Berger:2006db,Sadjadi:2006qp,Olivares:2007rt,
He:2009pd,Barreira:2011qi,delCampo:2008jx,Jamil:2009zzc,Jamil:2011iu}.

In this section, we intend to extend $(m,n)$-type holographic dark
energy models with interactions. When the interaction is taken into
account, the equations of motion for $\rho_{\Lambda}$ and $\rho_i$
become
\begin{equation}
\dot{\rho}_{\Lambda}=-3H\rho_{\Lambda}(1+\omega)-Q,\label{idd1}%
\end{equation}%
\begin{equation}
\dot{\rho}_{i}=-3H\rho_{i}(1+{\omega_{i}})+Q,\label{idd2}%
\end{equation}
where $Q$ denotes the interacting term. From (\ref{idd1}) and
(\ref{idd2}) we find that the interacting term has the following
general form,
\begin{equation}
\tilde{Q}\equiv{\frac{Q}{H\rho_{\Lambda}}}={\frac{1}{1+r}}[r^{\prime}-3(\omega-\omega_{i}) r].\label{h3}%
\end{equation}

It is also easy to derive a general relation between $L$ and
$\tilde{Q}$ as
\begin{equation}
\tilde{Q}=r^{\prime}-2r({\frac{L^{\prime}}{L}}-{\frac{3}{2}}-{\frac{3\omega_i}{2}}).\label{qo}%
\end{equation}
As we stressed in Ref.\cite{Hu:2006ar}, four free parameters
$\omega, r, L$ and $Q$ are not independent. Given any two of them,
the dynamics of the other two will be determined. Usually, people
propose the forms of $L$ and $Q$, and then find out the evolutions
of $\omega$ and $r$ with observation data. Thus, after introducing
the interacting term we find the equations for $\omega$ and $r$ in
the previous section can be generalized as
\begin{equation} \omega
=-1-{2\over 3}m+{2\over
3}\frac{a^{n-m}}{c\sqrt{1+r}}-{\tilde{Q}\over 3}.\label{es2}
\end{equation}
\begin{equation} r^{\prime}=\tilde{Q}(1+r)+3(\omega-\omega_i)r.\label{qw}
\end{equation}
Obviously the interaction will change the dynamics of $\omega$ as
well as $r$. One can alternatively write down the equation of
motion for $\Omega_{\Lambda}$ as
\begin{equation}{\Omega^{\prime}_{\Lambda}}={\Omega_{\Lambda}}[(3+3\omega_i+2m-{2a^{n-m}\sqrt{\Omega_{\Lambda}}\over{c}})(1-{\Omega_{\Lambda}})-{\tilde{Q}}{\Omega_{\Lambda}}].\label{w2}
 \end{equation}
 It is clear that Eqs.(\ref{es2}) and (\ref{w2}) reduce to Eqs.(\ref{es1}) and (\ref{w1}) respectively in the case of $\tilde{Q}=0$.
Now, we turn to consider the coincidence problem with the help of
interaction. We expect that the ratio $r$ of dark matter to dark
energy density varies slowly, and will finally approach to a
non-zero constant at late time. For explicitness, we consider a
specific form of the interaction $\tilde{Q}=3b^2(r+1)$, where
$b^2$ is a coupling constant. Its positivity is responsible for
the transition from dark energy to dark matter.
 Repeating the calculations in the previous section, we
find that the basic constraint on $m$ and $c$  becomes
\begin{equation}
m > -{5\over 6}-2b^2,
\end{equation}
\begin{equation}
c >{ 3\sqrt{3}\over 6m+5+12b^2}.
\end{equation}
To alleviate the coincidence problem, we are more concerned with
the asymptotic value of $r$ as $(a\gg 1)$. Setting $r^{\prime}=0$
in Eq.(\ref{qw}), we find the ratio of dark matter to dark energy
will approach to a non-zero constant, which is

\begin{equation}
 r_f={3b^2\over{3+2m-3b^2}},\label{rf}
\end{equation}
where the value of $r_f$ depends on $m$ and $b$ manifestly. This
result indicates that if $m$ is not too large, the situation that
the ratio $r$ keeps staying in a region with unit order can be
easily realized, thus providing a mechanism to understand the
coincidence problem.

\section{$(m,n)$ type models with a generalized future event horizon}
In this section, we would like to point out that with the same
spirit our construction should be applicable to the holographic dark
energy models with generalized future event horizon as the
characteristic size, which has been extensively studied in
literature\cite{Li:2011sd}. Explicitly, we may generalize the
definition of the holographic characteristic scale to
\begin{equation} L={1\over a^m(t)}\int^\infty_t a^n(t')dt'.\label{l1}
\end{equation}
Specially, when $(m,n)=(-1,-1)$ it recovers the ordinary holographic
dark energy model with future event horizon. In this definition
taking the derivative with respective to $lna$ on both sides, we
obtain
\begin{equation}
{L^{\prime}\over{L}}=-m-{a^{n-m}\over{HL}}.
\end{equation}
With the same algebra we may derive the equation of state as
\begin{equation}
\omega=-1-{2\over3}m-{2a^{n-m}\sqrt{\Omega_\Lambda}\over{3c}},\label{d2}
\end{equation}
while the equation of motion for $\Omega_\Lambda$ reads as
\begin{equation}
{\Omega^{\prime}_{\Lambda}}={\Omega_{\Lambda}}(1-{\Omega_{\Lambda}})(3+3\omega_i+2m+{2a^{n-m}\sqrt{\Omega_{\Lambda}}\over{c}}).\label{d3}
\end{equation}
In general case without interaction we still require that $n\leq
m$ such that the proportion of dark energy always increases with
the evolution of the universe if $m>-{\frac{3}{2}}$. Moreover,
under the condition of acceleration $1+r+3\omega<0$, and with the
use of $r_{0}\simeq 1/3$ we find the number $m$ should be subject
to the inequality
\begin{equation}m>-{5\over 6} -{\sqrt{3}\over{2c}}.
\end{equation}

It is interesting to notice that the fate of the universe would be
very different for $n=m$ and $n<m$ in future with $a\gg1$. For
$n<m$, we easily find that the asymptotic behavior of the state
parameter
 will be depicted by the equation
\begin{equation}\omega=-1-{2\over3}m,\label{st3}
\end{equation}
while for $m=n$, we find its value will approach to
\begin{equation}
\omega=-1-{2\over3}m-{2\over 3c},\label{st4}
\end{equation}
which depends on the constant $c$. The latter ones of course cover
the ordinary holographic dark energy model with $m=n=-1$ and
$c=1$. Based on our construction we could consider a generalized
model with $m=-1$ and $n=-1-\delta$ where $\delta$ is a small
positive constant. From our above consideration it is expected
that this modification will change the asymptotical behavior of
the dark energy dramatically. From this point our construction
here is quite different from the generalized holographic model
with varying $c(z)$, which has recently been proposed in
\cite{Zhang:2012qra}.

\section{the stability of $(m,n)$ type holographic dark energy model}

In above sections we have discussed the cosmic evolution of
our model. In this section we are concerned with the
stability of our model. Let us start with
 the scalar type perturbation of the metric in flat universe,
 which is written as
 \begin{equation}
 ds^2=-(1+2\Phi)dt^2+a^2(1-2\Phi)(dr^2+r^2d\Omega^2),
 \end{equation}
 where $\Phi$ is the Newtonian potential. Now  we can use the characteristic
scale without perturbation to define $r_{L0}(t)$ which is
\begin{equation}
L(0)\equiv L=a(t)r_{L0}(t).\label{lar}
\end{equation}
When the scalar perturbation is taken into account, the
characteristic scale is modified as,
\begin{equation}
L(\Phi)=a(t)\int^{r_{L}(t)}_0[1-\Phi(r',t)]dr'.
\end{equation}
 We consider the 00-component of the perturbation Einstein
equation in the Newtonian gauge. Making use of the variation of
Friedmann equation (\ref{f1}) and expanding $\Phi$ as
$\Phi=\sum\Phi_l{\sin lr\over r}$, we find it can be written as
\begin{equation}
-{\sin lr\over r}[{l^2\over
a^2}\Phi_l(t)+3H\dot{\Phi}_l(t)+3H^2\Phi_l(t)]
=-{3H\Phi_l(t)\dot{r}_{L0}\over r^2_{L0}}[\sin
lr_{L0}-\int^{r_{L0}}_0{\sin lr'\over r'}dr'].\label{ptur}
\end{equation}
Here we have used relations $\delta L(0)=L(\Phi)-L(0)$ and $\delta
L(0)=a(t)\delta r_{L0}+r_{L0}\delta a(t)$. Moreover, it is worth
to note that the above equation is independent of the specific
form of the cosmological model. To investigate the stability of
our model,  we are mainly concerned with the asymptotic behavior
of ${\dot{\Phi}_l\over\Phi_l}$ when $a\gg1$. As shown in
\cite{Li:2008zq,Huang:2013mqa}, there are two cases corresponding
to the stability: (1) the perturbation mode is frozen
when${\dot{\Phi}_l\over\Phi_l}\to 0$; (2) the perturbation mode is
decaying when ${\dot{\Phi}_l\over\Phi_l}<0$.

\subsection{the stability of the model with the age-like characteristic scale}
For the age-like holographic dark energy model, we define
the coordinate value by Eq.(\ref{s1}), which is
\begin{equation}
r_{L0}(t)={1\over a^{m+1}(t)}\int^t_0a^n(t')dt'.
\end{equation}
 Under the basic constraints $m>n$ and $m>-{5\over6}$, we will discuss the stability of the model for
  two cases  respectively,  namely $-{5\over6}<m<0$ and $m\ge0$.
 \begin{itemize}
\item When $-{5\over6}<m<0$, employing Eq.(\ref{st1}) we can
obtain the asymptotic behavior of dark energy density from the
continuity equation (\ref{i1}). It turns out that as $a\gg 1$,
$\rho_{\Lambda}\rightarrow 0$ and $L \to \infty$. For
super-horizon modes, namely $lr_{L0}\ll1$, we can derive the
following result from Eq.(\ref{ptur})
\begin{equation}
{\dot{\Phi}_l\over\Phi_l}=-{1\over3H}[{l^2\over a^2}+3H^2]
\label{ptu}.
\end{equation}
 Similarly, for sub-horizon mode $(lr_{L0}\gg1)$, we have
\begin{eqnarray}
{\dot{\Phi}_l\over\Phi_l}&=&-{1\over3H}[{l^2\over a^2}+3H^2]+{1\over L}{[a^{n-m}-(m+1)c][\sin lr_{L0}-{\pi\over2}]\over r_{L0}{\sin lr\over r}}\to0
\end{eqnarray}
 Above two equations indicate that our model is stable
when $-{5\over6}<m<0$.

\item  When $m\ge 0$, we notice that dark energy density
$\rho_{\Lambda}$ will approach to a constant ($m=0$) or
infinity($m>0$). For both cases we can derive $r_{L0}\to 0$ from
equation (\ref{lar}). As a result, for any given $l$, as $a\gg1$,
we always have $lr_{L0}\to 0$, implying that the sub-horizon modes
of the perturbation
 are absent. Thus we just need to consider the
super-horizon mode$(lr_{L0}\ll1)$, which gives rise to  the same
result as Eq.(\ref{ptu}).
\end{itemize}
As a summary, we conclude
that when the parameters $(m,n)$ are taken values in the allowed
region, we find our model is always stable under the scalar
perturbations.

\subsection{the stability of the model with the generalized event horizon as characteristic scale}

 We may consider the stability of the model with the
generalized event horizon as characteristic scale in a parallel
way. We define the coordinate $r_{L0}$  with the
generalized event horizon in equation (\ref{l1}), which is
\begin{equation}
r_{L0}={1\over a^{m+1}(t)}\int^\infty_t a^n(t')dt'.
\end{equation}
 Under the condition $m\geq-1$, we will discuss the stability for
 $m>n$ and $m=n$ respectively. For $m>n$, we can further classify them into two cases.
\begin{itemize}
\item In the case of $-1\leq m<0$, dark energy density
$\rho_{\Lambda}$ approaches to $0$ as $a\gg1$, which also implies
the generalized event horizon $L\to \infty$. For the super-horizon
modes we still have the same result as (\ref{ptu}). For
sub-horizon modes we can easily derive
\begin{eqnarray}
{\dot{\Phi}_l\over\Phi_l}&=&-{1\over3H}[{l^2\over a^2}+3H^2]-{1\over L}{[a^{n-m}+(m+1)c][\sin lr_{L0}-{\pi\over2}]\over r_{L0}{\sin lr\over r}} \to 0
\end{eqnarray}
\item In the case of $m\ge0$, $r_{L0}\to 0$ when $a\gg1$.
According to the previous analysis we can easily find that only
the super-horizon modes are presented and the model is stable too.
\end{itemize}
Next, we turn to analyze the perturbation behavior for the case
$m=n$. From equation(\ref{st4}), the future of our universe is
obviously related to the parameter $c$. Here we can choose a
specific value for $c$ to study the stability of our model.
Without loss of generality, we set $c=1$.

When $m\ge-1$, using equation (\ref{st4}) we find the perturbation
modes decay as described by equation (\ref{ptu}), which means our
model is stable. In particular, for $m=n=-1$, our discussion is
consistent with the analysis presented in \cite{Li:2008zq}, but
simpler. Our analysis above still holds for other values of
parameter $c$.

Therefore, there is no instability appearing for all the
perturbation modes in this sort of cosmological models.

\section{Discussion and Conclusions}

In this paper, we have constructed $(m,n)$ type holographic dark
energy model which can be viewed as a generalization of the ordinary
holographic dark energy models appeared in literature. At
phenomenological level, such a generalization provides us more space
in theory to fit the observational data. In particular, for some
specific values of $(m,n)$ the equation of state $\omega$ can
naturally evolve cross phantom divide in the entire evolution of the
universe even without introducing an interaction between dark energy
and background matter. We have discussed the general features of
age-like holographic dark energy models in various epoches of the
universe and derived the basic constraints on the values of $(m,n)$.
We have also remarked that this construction is applicable to the
holographic models with generalized future event horizon as the
characteristic scale.

For age-like holographic models, the case of $m=n$ is special. In
this case it seems that dark energy has the same behavior as the
dominant ingredient in the early epoches of the universe, implying
that dark energy might be unified with dark matter, analogous to
what happened in cosmological models with generalized Chaplygin
gas\cite{Bento:2002ps}. However, if DE and DM were unified at
early stages, we must introduce some mechanism to make dark energy
deviate from dark matter state, and eventually become dominant to
be responsible for the acceleration of the universe. This might be
implemented by introducing some appropriate interactions between
dark energy and dark matter, and our investigation is under
progress.

 We have investigated the stability issue by treating
holographic dark energy perturbation as global perturbation,
namely the perturbation of cosmic metric. In Newtonian gauge,
 we have shown that when the parameters are taken values in the allowed region,
 our model is always stable in the dark energy dominated era.

 We claim that this paper is our first step in this
direction, and our focus is proposing such an original model and
discussing its general features. We have found that our model
satisfies all the basic requirements to be a candidate of dark
energy. Further fitting with the observational data with more
quantitative precision of course is a key issue to testify our
models in the next step.

Note: After we uploaded our manuscript, the observational
constraint on this model appeared in \cite{Huang:2012xm}, in which
some integer parameters $(m,n)$ were analyzed in great detail. The
best-fit analysis in this reference indicates that this model with
$m = n + 1$ and small $m$ is more favored including the cases of
(m,n)=(1,0), (2,1), (3,2). In a word, the preliminary numerical
analysis has indicated that our model can fit with the
observational data very well.

\section*{Acknowledgement} We are grateful to Rong-gen Cai,
Hao Wei, and Yue-Liang Wu for helpful discussions and
correspondence. We are also grateful to the anonymous
referee for helpful suggestions. This work is partly supported by
NSFC (10875057,11178002), Fok Ying Tung Education Foundation
(No.111008), the key project of Chinese Ministry of Education
(No.208072), Jiangxi young scientists (JingGang Star) program and
555 talent project of Jiangxi Province.


\begin{thebibliography}{99}
\bibitem{Riess:1998cb}
  A.~G.~Riess {\it et al.}  [Supernova Search Team Collaboration],
  Astron.\ J.\  {\bf 116}, 1009 (1998)  [astro-ph/9805201].  
\bibitem{Perlmutter:1998np}
  S.~Perlmutter {\it et al.}  [Supernova Cosmology Project Collaboration],
  Astrophys.\ J.\  {\bf 517}, 565 (1999)  [astro-ph/9812133].  
\bibitem{Spergel:2003cb}
  D.~N.~Spergel {\it et al.}  [WMAP Collaboration],
  Astrophys.\ J.\ Suppl.\  {\bf 148}, 175 (2003)  [astro-ph/0302209].  
\bibitem{Tegmark:2003ud}
  M.~Tegmark {\it et al.}  [SDSS Collaboration],
  Phys.\ Rev.\ D {\bf 69}, 103501 (2004)  [astro-ph/0310723].  
\bibitem{Weinberg:1988cp}
  S.~Weinberg,
  Rev.\ Mod.\ Phys.\  {\bf 61}, 1 (1989).  
\bibitem{Weinberg:2000yb}
  S.~Weinberg,
  astro-ph/0005265.
\bibitem{Copeland:2006wr}
  E.~J.~Copeland, M.~Sami and S.~Tsujikawa,
  Int.\ J.\ Mod.\ Phys.\ D {\bf 15}, 1753 (2006)  [hep-th/0603057].  
\bibitem{Li:2011sd}
  M.~Li, X.~-D.~Li, S.~Wang and Y.~Wang,
  Commun.\ Theor.\ Phys.\  {\bf 56}, 525 (2011)  [arXiv:1103.5870 [astro-ph.CO]].  
\bibitem{'tHooft:1993gx}
  G.~'t Hooft,
  gr-qc/9310026.
\bibitem{'tHooft:1999bw}
  G.~'t Hooft,
  hep-th/0003004.
\bibitem{Susskind:1994vu}
  L.~Susskind,
  J.\ Math.\ Phys.\  {\bf 36}, 6377 (1995)  [hep-th/9409089].  
\bibitem{Cohen:1998zx}
  A.~G.~Cohen, D.~B.~Kaplan and A.~E.~Nelson,
  Phys.\ Rev.\ Lett.\  {\bf 82} (1999) 4971  [hep-th/9803132].  
\bibitem{Horava:2000tb}
  P.~Horava and D.~Minic,
  Phys.\ Rev.\ Lett.\  {\bf 85}, 1610 (2000)
  [hep-th/0001145].
\bibitem{Thomas:2002pq}
  S.~D.~Thomas,
  Phys.\ Rev.\ Lett.\  {\bf 89}, 081301 (2002).
\bibitem{Fischler:1998st}
  W.~Fischler and L.~Susskind,
  hep-th/9806039.
\bibitem{Bousso:1999xy}
  R.~Bousso,
  JHEP {\bf 9907}, 004 (1999)
  [hep-th/9905177].
\bibitem{Hsu:2004ri}
  S.~D.~H.~Hsu,
  Phys.\ Lett.\ B {\bf 594}, 13 (2004)
  [hep-th/0403052].
\bibitem{Li:2004rb}
  M.~Li,
  Phys.\ Lett.\ B {\bf 603}, 1 (2004)  [hep-th/0403127].  
\bibitem{Cai:2007us}
  R.~-G.~Cai,
  Phys.\ Lett.\ B {\bf 657}, 228 (2007)  [arXiv:0707.4049 [hep-th]].  
\bibitem{Wei:2007ty}
  H.~Wei and R.~-G.~Cai,
  Phys.\ Lett.\ B {\bf 660}, 113 (2008)  [arXiv:0708.0884 [astro-ph]].  
\bibitem{Gao:2007ep}
  C.~Gao, F. Wu, X.~Chen and Y.~-G.~Shen,
  Phys.\ Rev.\ D {\bf 79}, 043511 (2009)  [arXiv:0712.1394 [astro-ph]].  

\bibitem{Huang:2012nz}
 Z.~-P.~Huang and Y.~-L.~Wu,
  Int.\ J.\ Mod.\ Phys.\ A {\bf 27}, 1250085 (2012)
  [arXiv:1202.2590 [hep-th]].

\bibitem{Wei:2012wt}
  H.~Wei,
  Class.\ Quant.\ Grav.\  {\bf 29}, 175008 (2012)
  [arXiv:1204.4032 [gr-qc]].


\bibitem{Myung:2007pn}
  Y.~S.~Myung,
  Phys.\ Lett.\ B {\bf 652}, 223 (2007)  [arXiv:0706.3757 [gr-qc]].
\bibitem{Kim:2007iv}
  K.~Y.~Kim, H.~W.~Lee and Y.~S.~Myung,
  Phys.\ Lett.\ B {\bf 660}, 118 (2008)  [arXiv:0709.2743 [gr-qc]].
\bibitem{Li:2008zq}
  M.~Li, C.~Lin and Y.~Wang,
   JCAP {\bf 0805}, 023 (2008)  [arXiv:0801.1407 [astro-ph]].

\bibitem{Wei:2007xu}
  H.~Wei and R.~G.~Cai,
  Phys.\ Lett.\  B {\bf 663}, 1 (2008)
  [arXiv:0708.1894 [astro-ph]].

\bibitem{Huang:2012gd}
  Z.~-P.~Huang and Y.~-L.~Wu,
  Int.\ J.\ Mod.\ Phys.\ A {\bf 27}, 1250130 (2012)
  [arXiv:1202.3517 [astro-ph.CO]].



\bibitem{Hu:2006ar}
  B.~Hu and Y.~Ling,
  Phys.\ Rev.\  D {\bf 73}, 123510 (2006)
  [arXiv:hep-th/0601093].
\bibitem{Amendola:1999er}
  L.~Amendola,
  Phys.\ Rev.\ D {\bf 62}, 043511 (2000)
  [astro-ph/9908023].
\bibitem{Zimdahl:2001ar}
  W.~Zimdahl and D.~Pavon,
  Phys.\ Lett.\ B {\bf 521}, 133 (2001)
  [astro-ph/0105479].
\bibitem{Farrar:2003uw}
  G.~R.~Farrar and P.~J.~E.~Peebles,
  Astrophys.\ J.\  {\bf 604}, 1 (2004)
  [astro-ph/0307316].
\bibitem{Gumjudpai:2005ry}
  B.~Gumjudpai, T.~Naskar, M.~Sami and S.~Tsujikawa,
  JCAP {\bf 0506}, 007 (2005)  [hep-th/0502191].
\bibitem{CalderaCabral:2008bx}
  G.~Caldera-Cabral, R.~Maartens and L.~A.~Urena-Lopez,
  Phys.\ Rev.\ D {\bf 79}, 063518 (2009)  [arXiv:0812.1827 [gr-qc]].
\bibitem{Wu:2008jt}
  J.~-P.~Wu, D.~-Z.~Ma and Y.~Ling,
   Phys.\ Lett.\ B {\bf 663}, 152 (2008)  [arXiv:0805.0546 [hep-th]].

\bibitem{Sheykhi:2010nv}
  A.~Sheykhi and M.~Jamil,
  Phys.\ Lett.\ B {\bf 694}, 284 (2011)  [arXiv:1010.0385 [hep-th]].


\bibitem{Jamil:2010xq}
  M.~Jamil and M.~U.~Farooq,
  JCAP {\bf 1003}, 001 (2010)  [arXiv:1002.1434 [gr-qc]].

\bibitem{Avelino:2012tc}
  P.~P.~Avelino and H.~M.~R.~da Silva,
  Phys.\ Lett.\ B {\bf 714}, 6 (2012)
  [arXiv:1201.0550 [astro-ph.CO]].


\bibitem{Setare:2006wh}
   M RSetare,
  Phys.\ Lett.\ B {\bf 642}, 1 (2006)  [hep-th/0609069].
\bibitem{Setare:2007we}
  M.~R.~Setare and E.~C.~Vagenas,
  Int.\ J.\ Mod.\ Phys.\ D {\bf 18}, 147 (2009)  [arXiv:0704.2070 [hep-th]].

\bibitem{Guo:2007zk}
  Z.~-K.~Guo, N.~Ohta and S.~Tsujikawa,
  Phys.\ Rev.\ D {\bf 76}, 023508 (2007)  [astro-ph/0702015 [ASTRO-PH]].
\bibitem{He:2008tn}
  J.~-H.~He and B.~Wang,
  JCAP {\bf 0806} (2008) 010  [arXiv:0801.4233 [astro-ph]].
\bibitem{Feng:2008fx}
  C.~Feng, B.~Wang, E.~Abdalla and R.~-K.~Su,
  Phys.\ Lett.\ B {\bf 665}, 111 (2008)  [arXiv:0804.0110 [astro-ph]].

\bibitem{TocchiniValentini:2001ty}
  D.~Tocchini-Valentini and L.~Amendola,
  Phys.\ Rev.\ D {\bf 65}, 063508 (2002)  [astro-ph/0108143].
\bibitem{Cai:2004dk}
  R.~-G.~Cai and A.~Wang,
  JCAP {\bf 0503}, 002 (2005)  [hep-th/0411025].
\bibitem{Berger:2006db}
  M.~S.~Berger and H.~Shojaei,
  Phys.\ Rev.\ D {\bf 73}, 083528 (2006)  [gr-qc/0601086].
\bibitem{Sadjadi:2006qp}
  H.~M.~Sadjadi and M.~Alimohammadi,
  Phys.\ Rev.\ D {\bf 74}, 103007 (2006)  [gr-qc/0610080].
\bibitem{Olivares:2007rt}
  G.~Olivares, F.~Atrio-Barandela and D.~Pavon,
  Phys.\ Rev.\ D {\bf 77}, 063513 (2008)  [arXiv:0706.3860 [astro-ph]].
\bibitem{He:2009pd}
  J.~-H.~He, B.~Wang and P.~Zhang,
  Phys.\ Rev.\ D {\bf 80}, 063530 (2009)  [arXiv:0906.0677 [gr-qc]].

\bibitem{Barreira:2011qi}
  A.~Barreira and P.~P.~Avelino,
  Phys.\ Rev.\ D {\bf 83}, 103001 (2011)  [arXiv:1103.2401 [astro-ph.CO]].
\bibitem{delCampo:2008jx}
  S.~del Campo, R.~Herrera and D.~Pavon,
  JCAP {\bf 0901}, 020 (2009)  [arXiv:0812.2210 [gr-qc]].
\bibitem{Jamil:2009zzc}
  M.~Jamil and F.~Rahaman,
  Eur.\ Phys.\ J.\ C {\bf 64}, 97 (2009).
\bibitem{Jamil:2011iu}
  M.~Jamil, D.~Momeni and M.~A.~Rashid,
  Eur.\ Phys.\ J.\ C {\bf 71}, 1711 (2011)  [arXiv:1107.1558 [physics.gen-ph]].

\bibitem{Zhang:2012qra}
  Z.~Zhang, M.~Li, X.~-D.~Li, S.~Wang and W.~-S.~Zhang,
  Mod.\ Phys.\ Lett.\ A {\bf 27}, 1250115 (2012)
  [arXiv:1202.5163 [astro-ph.CO]].

 \bibitem{Huang:2013mqa}
  P.~Huang and Y.~-c.~Huang,
  Eur.\ Phys.\ J.\ C {\bf 73}, 2366 (2013)
  [arXiv:1304.0366 [hep-th]].

\bibitem{Bento:2002ps}
  M.~C.~Bento, O.~Bertolami and A.~A.~Sen,
  Phys.\ Rev.\ D {\bf 66}, 043507 (2002)  [gr-qc/0202064].

\bibitem{Huang:2012xm}
  Z.~-P.~Huang and Y.~-L.~Wu,
  JCAP {\bf 1207}, 035 (2012)
  [arXiv:1205.0608 [gr-qc]].







\end{thebibliography}
\end{document}